\documentclass[journal]{IEEEtran}
\usepackage{cite}
\usepackage{amsmath,amssymb,amsfonts}
\usepackage{algorithmic}
\usepackage{algorithm}
\usepackage{graphicx}
\usepackage{textcomp}
\usepackage{xcolor}

\usepackage[
  colorlinks=true,
  linkcolor=blue,
  citecolor=blue,
  urlcolor=blue
]{hyperref}

\begin{document}
%
\title{Non-Reciprocal Dynamic Metasurface Antenna: Practical Multiport-Network Modeling and Optimization for Multi-User Interference Resilience}
%
%
%

\author{Shuai S. A. Yuan,~\IEEEmembership{Member,~IEEE,}
        Jean Tapie,~\IEEEmembership{Graduate Student Member,~IEEE,}
        Bahman Amrahi, \\
        Viktar Asadchy,~\IEEEmembership{Senior Member,~IEEE,}
        and Philipp del Hougne,~\IEEEmembership{Member,~IEEE}

\thanks{
S.~S.~A.~Yuan, B.~Amrahi, V.~Asadchy, and P.~del Hougne are with the Department of Electronics and Nanoengineering, Aalto University, 00076 Espoo, Finland. J.~Tapie and P.~del~Hougne are with Univ Rennes, CNRS, IETR - UMR 6164, F-35000, Rennes, France. (e-mail: shuai.yuan@aalto.fi; jean-tapie@univ-rennes.fr; bahman.amrahi@aalto.fi; viktar.asadchy@aalto.fi; philipp.del-hougne@univ-rennes.fr)
}
\thanks{\textit{(Corresponding Author: Philipp del Hougne.)}}
\thanks{This work was supported in part by the Nokia Foundation (project 20260028), the ANR France 2030 program (project ANR-22-PEFT-0005), the ANR PRCI program (project ANR-22-CE93-0010), the French Defense Innovation Agency (project 2024600), the European Union's European Regional Development Fund, the French region of Brittany and Rennes Métropole through the contrats de plan État-Région program (projects ``SOPHIE/STIC \& Ondes'' and ``CyMoCoD''), and the Research Council of Finland (projects 371367 and 365679).}

}


\maketitle

\begin{abstract}
Channel reciprocity fundamentally limits full-duplex (FD) base stations due to multi-user co-channel interference. We examine the potential of deploying a non-reciprocal dynamic metasurface antenna (NR-DMA) at the base station to overcome this limitation. Our NR-DMA architecture connects a single circulator to three feed ports of a multi-feed DMA with strong mutual coupling (MC) between its seven feeds and 96 1-bit-programmable meta-elements. We model our system with multiport network theory, using experimentally estimated proxy parameters of a fabricated 19-GHz DMA and the measured circulator response. Our NR-DMA's reconfigurability is captured by a \textit{diagonal} tunable scattering matrix, showing that non-reciprocal DMAs and RISs need not require a ``beyond-diagonal'' tunable scattering matrix. We jointly optimize the DMA state, analog feed weights, circulator-port assignment, and circulation direction. Our optimized NR-DMA realizes distinct forward and reverse channel responses. In our interference-limited high-SNR case study, the NR-DMA improves the FD sum rate by about \(60\%\) over a reciprocal DMA benchmark. Comparisons with proxy objectives and MC-unaware optimization show that end-to-end FD optimization and MC-aware modeling are both essential.
\end{abstract}

\begin{IEEEkeywords}
Dynamic metasurface antenna, full-duplex communications, multiport network theory, multi-user interference, mutual coupling, non-reciprocal beamforming, non-reciprocity.
\end{IEEEkeywords}

\section{Introduction}

Managing interference between desired and undesired wireless signals is notoriously challenging in advanced wireless networks with full-duplex (FD) entities that simultaneously transmit and receive. Channel reciprocity usually implies that a propagation configuration enhancing the downlink (DL) channel from a base station (BS) toward one terminal also enhances the corresponding reverse uplink (UL) channel from the same terminal direction back to the BS. Hence, in a FD multi-user setting, a BS that radiates efficiently toward a DL user also presents high receive sensitivity to co-channel UL emissions arriving from that same direction, even when the desired UL signal originates from a different user.

A potential remedy lies in incorporating non-reciprocity into the BS. While the wireless propagation environment itself is typically reciprocal, the BS can include non-reciprocal components that decouple its transmit and receive radiation patterns. In this Letter, we explore the performance gains of this concept using a dynamic metasurface antenna (DMA) as the BS aperture. DMAs are a promising hybrid analog/digital beamforming technology distinguished by low power consumption, small footprint, and low cost~\cite{shlezinger2021dynamic}. To the best of our knowledge, all existing DMA architectures are fundamentally reciprocal. Inspired by~\cite{yuan2026single}, we convert an existing reciprocal multi-feed DMA into a non-reciprocal DMA (NR-DMA) by connecting a single circulator to three DMA feeds. 
The efficacy of our approach is enabled by strong coupling between all feeds and all meta-elements in our DMA, mediated by a chaotic-cavity-based structure. The same mechanism by which this strong all-to-all coupling boosts the DMA's wave-domain flexibility~\cite{prod2025mutual,prod2025benefits} also amplifies the system-level impact of the circulator-induced non-reciprocity.

Conceptually related to NR-DMAs are non-reciprocal reconfigurable intelligent surfaces (NR-RISs). The main difference between a DMA and an RIS is that feeds are integrated into the DMA, whereas an RIS requires separate feed antennas, resulting in a bulkier setup. The physics-consistent system models for DMAs and RISs based on multiport-network theory (MNT) are essentially identical, treating tunable lumped elements as ``virtual'' ports terminated by tunable loads; see Sec.~\ref{sec_SystemModel}. 
NR-RISs have so far mainly been studied using non-diagonal scattering matrices representing non-reciprocal load networks to terminate these ``virtual'' ports, for applications including channel reciprocity attacks~\cite{wang2024channel,xu2025non} and FD communications~\cite{li2025non,liu2026non}.
Abstracting the load network as being able to realize, e.g., an arbitrary (unitary) scattering matrix neglects practical hardware impairments like loss and delay in static connections between the tunable loads and the quantized programmability of practical tunable loads. Moreover, this abstracted approach has led to the conclusion that NR-RISs are fundamentally beyond-diagonal RISs (BD-RISs)~\cite{li2025non,liu2026non}, i.e., RISs in which the ``virtual'' ports are not terminated by independent tunable loads but by a tunable load network with tunable connections between ``virtual'' ports. 

While a diagonal load matrix (describing an ensemble of individual tunable loads) is symmetric and hence unable to introduce non-reciprocity, the non-reciprocity may be incorporated into the static RIS or DMA structure rather than into the tunable load network. Moreover, practical realizations of non-reciprocal tunable load networks typically consist of static non-reciprocal components (e.g., isolators, circulators, or gyrators) and individually tunable loads~\cite{xu2025non}. Consequently, the non-reciprocal components can be viewed as part of the static RIS or DMA structure in these cases and there exists an MNT representation involving a tunable \textit{diagonal} load matrix, in analogy with recent physics-consistent representations of BD-RISs~\cite{del2025physics} and BD-DMAs~\cite{prod2025beyond} based on tunable \textit{diagonal} load matrices. The representation involving a diagonal load matrix has the practical advantage of directly exposing the optimizable variables (the individual loads)~\cite{del2025physics,prod2025beyond}. 

\textit{Contributions:}
\textit{First}, we introduce the first non-reciprocal DMA architecture. By leveraging strong all-to-all coupling between feeds and meta-elements, our architecture uses a single circulator. This feature matters since non-reciprocal components are costly. In contrast, the number of non-reciprocal components in the NR-RIS in~\cite{xu2025non} scales with the number of ``virtual'' ports;~\cite{wang2024channel,li2025non,liu2026non} do not specify a concrete circuit to realize the abstracted non-reciprocal load network. 
\textit{Second}, we develop an experimentally grounded practical multiport-network model for our NR-DMA, based on experimentally estimated proxy MNT parameters for a fabricated chaotic-cavity-backed multi-feed DMA with 96 1-bit-programmable meta-elements and an experimentally characterized circulator.
\textit{Third}, we study the performance of our NR-DMA as an FD BS in a multi-user setting. We compare the achieved sum-rate with beamforming-based, channel-based, and rate-based optimization metrics.
\textit{Fourth}, we benchmark the achieved performance against a reciprocal DMA and against optimizations based on widespread coupling-unaware simplified system models.

\textit{Notation:}
$\mathbf{I}_a$ is the $a\times a$ identity matrix.
$\mathbf 0$ and $\mathbf 1$ denote, respectively, the all-zeros and the all-ones vectors/matrices of appropriate sizes.
$\mathrm{diag}(\mathbf{a})$ denotes the diagonal matrix whose diagonal entries are given by the vector $\mathbf{a}$.
$\mathbf{A}_{\mathcal{B}\mathcal{C}}$ denotes the block of $\mathbf{A}$ selected by row indices $\mathcal{B}$ and column indices $\mathcal{C}$.

\section{System model}
\label{sec_SystemModel}

\begin{figure}
    \centering
    \includegraphics[width=0.85\linewidth]{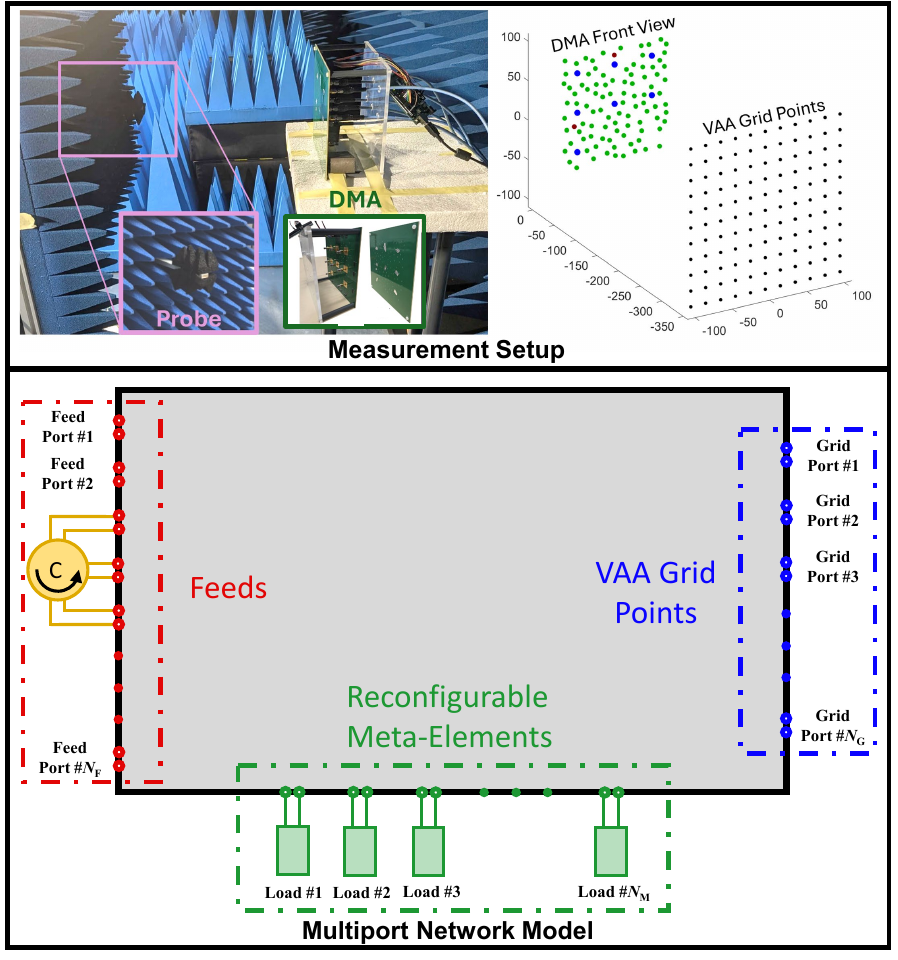}
    \caption{Top: VAA-based measurement setup for the reciprocal multi-feed DMA. Bottom: MNT-based system model for our NR-DMA.}
    \label{fig:multiport_model}
\end{figure}

\textit{Multi-Feed Reciprocal DMA:} Following~\cite{tapie2026experimental}, we briefly summarize the MNT-based system model for a DMA together with a virtual
antenna array (VAA) used to sample the DMA's radiated field. This model applies to any DMA architecture whose reconfigurability is based on tunable lumped elements, and it makes no assumptions about the DMA-VAA separation or the VAA geometry. The DMA is abstracted as a radiating structure with $N_\mathrm{F}$ lumped feed ports and $N_\mathrm{M}$ tunable lumped elements; the DMA's radiation pattern is sampled by the VAA at $N_\mathrm{G}$ grid points. Each of the DMA's $N_\mathrm{M}$ tunable lumped elements is represented as a ``virtual'' port terminated by a tunable load. The system comprising DMA and VAA can be partitioned into an $N$-port static subsystem, where $N=N_\mathrm{F}+N_\mathrm{M}+N_\mathrm{G}$, and an $N_\mathrm{M}$-port tunable subsystem, connected via the $N_\mathrm{M}$ ``virtual'' ports. The static subsystem is characterized by its scattering matrix $\mathbf{S}\in\mathbb{C}^{N \times N}$ and the tunable subsystem is characterized by its scattering matrix $\mathbf{\Phi}=\mathrm{diag}(\mathbf{r})\in\mathbb{C}^{N_\mathrm{M} \times N_\mathrm{M}}$, where $\mathbf{r} = \left[r_1, r_2, \dots, r_{N_\mathrm{M}} \right]\in\mathbb{C}^{N_\mathrm{M}}$ is the DMA's load vector with $r_i\in\mathbb{C}$ being the reflection coefficient of the $i$th tunable load. Standard MNT yields the ``forward'' transmission matrix $\mathbf{T}_\mathrm{fwd}\in\mathbb{C}^{N_\mathrm{G}\times N_\mathrm{F}}$ from the DMA feed ports to the VAA:
\begin{equation}
\mathbf T_\mathrm{fwd}(\mathbf r)
=
\mathbf S_{\mathcal G\mathcal F}
+
\mathbf S_{\mathcal G\mathcal M}
\bigl(
\mathbf I_{N_\mathrm M}
-
\mathbf\Phi(\mathbf r)\mathbf S_{\mathcal M\mathcal M}
\bigr)^{-1}
\mathbf\Phi(\mathbf r)
\mathbf S_{\mathcal M\mathcal F},
\label{eq_DMA_MNT_T}
\end{equation}
where $\mathcal F$, $\mathcal M$, and $\mathcal G$ denote the port-index sets associated with the feeds, tunable elements, and VAA ports, respectively. Reciprocity implies $\mathbf{S}=\mathbf{S}^\top$ as well as $\mathbf{T}_\mathrm{rev}=\mathbf{T}_\mathrm{fwd}^\top$, where $\mathbf{T}_\mathrm{rev}\in\mathbb{C}^{N_\mathrm{F}\times N_\mathrm{G}}$ is the ``reverse'' transmission matrix from the VAA to the DMA feed ports. 

For practical DMAs, the possible values of $r_i$ are quantized, with both phase and amplitude depending on the element state. Oftentimes, including the case of our fabricated prototype, the tunable elements are PIN diodes such that there are only two possible states, i.e., $r_i \in \{\alpha,\beta\}$, and the mapping from the DMA's control vector $\mathbf{v}\in\{0,1\}^{N_\mathrm{M}}$ to its load vector $\mathbf{r}$ is affine:
\begin{equation}
\mathbf r(\mathbf v)
=
\alpha\mathbf 1
+
(\beta-\alpha)\mathbf v.
\label{eq_DMA_encoding}
\end{equation}

\textit{Extension to NR-DMA:} 
Assuming $N_\mathrm{F}>3$ (our prototype has $N_\mathrm{F}=7$), we partition the DMA feed ports into a set of three feeds that are terminated by a circulator and the remaining $N_\mathrm{F}-3$ feeds; the corresponding port index sets are $\mathcal{C}$ and $\mathcal{A}=\mathcal{F}\setminus\mathcal{C}$. The circulator is a three-port device characterized by its scattering matrix $\mathbf{S}^\mathrm{C}\in\mathbb{C}^{3\times3}$. The resulting MNT model is shown in the lower panel in Fig.~\ref{fig:multiport_model}. Because the termination of the ports in $\mathcal{C}$ by the circulator is fixed and independent of the DMA configuration, we can absorb it into a reduced static $\bar{N}$-port subsystem with only the remaining $N_\mathrm{F}-3$ feed ports, the $N_\mathrm M$ ``virtual'' ports, and the $N_\mathrm G$ VAA ports, i.e., $\bar{N} = (N_\mathrm{F}-3)+N_\mathrm{M}+N_\mathrm{G}$. The reduced static subsystem's scattering matrix $\bar{\mathbf{S}}\in\mathbb{C}^{\bar{N}\times\bar{N}}$ is given by standard MNT:
\begin{equation}
\bar{\mathbf S}
\triangleq
\mathbf S_{{\mathcal P}{\mathcal P}}
+
\mathbf S_{{\mathcal P}\mathcal C}
\left(
\mathbf I_{3}
-
\mathbf{S}^\mathrm{C}\mathbf S_{\mathcal C\mathcal C}
\right)^{-1} \mathbf{S}^\mathrm{C}
\mathbf S_{\mathcal C{\mathcal P}},
\label{eq_open_circuit_static_reduction}
\end{equation}
where $\mathcal{P}=\mathcal{A}\cup\mathcal{M}\cup\mathcal{G}$. Consequently, the counterpart of \eqref{eq_DMA_MNT_T} using this reduced static subsystem is
\begin{equation}
\bar{\mathbf T}_\mathrm{fwd}(\mathbf r)
=
\bar{\mathbf S}_{\mathcal G\mathcal A}
+
\bar{\mathbf S}_{\mathcal G\mathcal M}
\bigl(
\mathbf I_{N_\mathrm M}
-
\mathbf\Phi(\mathbf r)\bar{\mathbf S}_{\mathcal M\mathcal M}
\bigr)^{-1}
\mathbf\Phi(\mathbf r)
\bar{\mathbf S}_{\mathcal M\mathcal A},
\label{eq_DMA_MNT_reduced_T}
\end{equation}
where
$\bar{\mathbf T}_\mathrm{fwd}(\mathbf r)\in\mathbb C^{N_\mathrm G\times (N_\mathrm{F}-3)}$ is the transmission matrix from the remaining $N_\mathrm{F}-3$ DMA feeds to the VAA.
Analogously, the reverse transmission matrix $\bar{\mathbf{T}}_\mathrm{rev}\in\mathbb{C}^{(N_\mathrm{F}-3)\times N_\mathrm{G}}$ from the VAA to the remaining $N_\mathrm{F}-3$ DMA feed ports is
\begin{equation}
\bar{\mathbf T}_\mathrm{rev}(\mathbf r)
=
\bar{\mathbf S}_{\mathcal A\mathcal G}
+
\bar{\mathbf S}_{\mathcal A\mathcal M}
\bigl(
\mathbf I_{N_\mathrm M}
-
\mathbf\Phi(\mathbf r)\bar{\mathbf S}_{\mathcal M\mathcal M}
\bigr)^{-1}
\mathbf\Phi(\mathbf r)
\bar{\mathbf S}_{\mathcal M\mathcal G}.
\label{eq_DMA_MNT_reduced_Trev}
\end{equation}
Because the reduced static subsystem is non-reciprocal, generally $\bar{\mathbf{S}}\neq\bar{\mathbf{S}}^\top$ such that $\bar{\mathbf{T}}_\mathrm{rev}\neq\bar{\mathbf{T}}_\mathrm{fwd}^\top$, which is precisely the property we seek to decouple UL and DL channels.

\section{Physics-Consistent Practical NR-DMA Optimization}

\subsection{Problem Formulation}

\textit{Scenario:} We consider the NR-DMA as the aperture of an FD BS, where DL transmission and UL reception occur simultaneously over the same frequency band. The BS transmits a DL data stream toward a single-antenna DL terminal at VAA grid point \(g_\mathrm{DL}\), while simultaneously receiving a desired UL data stream from a different single-antenna UL terminal at grid point \(g_\mathrm{UL}\). In addition, an undesired co-channel UL signal impinges on the BS from grid point \(g_\mathrm{DL}\). The BS should thus radiate efficiently toward the DL terminal while suppressing receive sensitivity from the same direction, an aperture-level objective that naturally calls for non-reciprocity. This setting can represent either an interfering transmitter aligned with the DL terminal or an FD terminal with ideal terminal-side self-interference cancellation that simultaneously receives DL and emits UL. To focus on channel non-reciprocity, we assume ideal cancellation of the BS's self-interference and the absence of user-to-user interference (blocked user-to-user line-of-sight and negligible user--BS--user scattering paths).

\textit{Optimization Variables:}
For a given set of user locations, we jointly optimize four sets of variables:
\begin{itemize}
    \item the binary \textit{DMA control vector}
    \(\mathbf v\in\{0,1\}^{N_\mathrm M}\), which determines the DMA's load vector \(\mathbf r(\mathbf v)\) via~\eqref{eq_DMA_encoding};
    \item the \textit{common analog feed-weight vector}
    \(\mathbf x\in\mathbb C^{N_\mathrm F-3}\), with
    \(\|\mathbf x\|_2^2=1\), which models a shared analog feed network acting as the transmit precoder and receive combiner at the \(N_\mathrm{F}-3\) externally accessible DMA feeds;
    \item the \textit{circulator-port assignment}
    \(\boldsymbol\pi\in\Pi\), where each \(\boldsymbol\pi\) specifies which three DMA feed ports are connected to the three circulator terminals and in which order;
    \item the \textit{circulation direction} \(d\in\mathcal D\triangleq\{+,-\}\), corresponding to circulator orientation with $\mathbf{S}_\circlearrowleft =
\mathbf{S}_\circlearrowright^\top\triangleq\mathbf{S}^\mathrm{C}$.
\end{itemize}
For a given pair \((\boldsymbol\pi,d)\), the circulator-connected feed ports are absorbed into the reduced static $\bar{\mathbf{S}}^{(\boldsymbol\pi,d)}$ following \eqref{eq_open_circuit_static_reduction}.

\textit{Optimization Objectives:}
Let \(\mathbf e_g\) denote the canonical vector selecting VAA grid point \(g\). For a given \((\mathbf v,\boldsymbol\pi,d)\), the reduced MNT model yields the forward and reverse transmission matrices
\(\bar{\mathbf T}_\mathrm{fwd}^{(\boldsymbol\pi,d)}(\mathbf r(\mathbf v))\) and
\(\bar{\mathbf T}_\mathrm{rev}^{(\boldsymbol\pi,d)}(\mathbf r(\mathbf v))\). The corresponding channel gains between the BS and a terminal at grid point $g$ under the common analog feed-weight vector \(\mathbf x\) are
\begin{subequations}
\label{eq:gains}
\begin{align}
G_\mathrm{fwd}(g;\mathbf v,\mathbf x,\boldsymbol\pi,d)
&=
\left|
\mathbf e_g^\top
\bar{\mathbf T}_\mathrm{fwd}^{(\boldsymbol\pi,d)}(\mathbf r(\mathbf v))
\mathbf x
\right|^2,
\label{eq:gain_fwd}\\
G_\mathrm{rev}(g;\mathbf v,\mathbf x,\boldsymbol\pi,d)
&=
\left|
\mathbf x^\top
\bar{\mathbf T}_\mathrm{rev}^{(\boldsymbol\pi,d)}(\mathbf r(\mathbf v))
\mathbf e_g
\right|^2.
\label{eq:gain_rev}
\end{align}
\end{subequations}
For compactness, we abbreviate these quantities as
\(G_\mathrm{fwd}(g)\) and \(G_\mathrm{rev}(g)\), respectively, in the following.

Our main objective is to maximize the FD sum rate defined in~\eqref{eq:O_FD}; we additionally consider two proxy objectives.
\begin{itemize}
    \item \textit{NR Beamforming Proxy Objective:}
    maximize target-to-background contrast in the transmit and receive radiation maps,
    \begin{equation}
    \begin{aligned}
    \mathcal O_\mathrm{BF}
    =
    &\log
    \frac{G_\mathrm{fwd}(g_\mathrm{DL})+\varepsilon}
    {\max_{g\in\mathcal G\setminus\{g_\mathrm{DL}\}}G_\mathrm{fwd}(g)+\varepsilon}
    \\
    \quad+
    &\log
    \frac{G_\mathrm{rev}(g_\mathrm{UL})+\varepsilon}
    {\max_{g\in\mathcal G\setminus\{g_\mathrm{UL}\}}G_\mathrm{rev}(g)+\varepsilon}.
    \end{aligned}
    \label{eq:O_BF}
    \end{equation}
    The logarithms turn gain ratios into additive contrast terms, and \(\varepsilon>0\) avoids singular ratios.

    \item \textit{NR Channel Shaping Proxy Objective:}
    maximize dual-user channel contrast, i.e., desired over unintended responses at the DL and UL user locations,
    \begin{equation}
    \begin{aligned}
    \mathcal O_\mathrm{CH}
    &=
    \log
    \frac{G_\mathrm{fwd}(g_\mathrm{DL})+\varepsilon}
    {G_\mathrm{fwd}(g_\mathrm{UL})+\varepsilon}
    +
    \log
    \frac{G_\mathrm{rev}(g_\mathrm{UL})+\varepsilon}
    {G_\mathrm{rev}(g_\mathrm{DL})+\varepsilon}.
    \end{aligned}
    \label{eq:O_CH}
    \end{equation}
    Unlike \(\mathcal O_\mathrm{BF}\), this proxy only compares the two relevant user directions, while ignoring responses at other grid points.

    \item \textit{FD Sum-Rate Maximization Objective:}
    maximize the FD sum rate,
    \begin{equation}
    \begin{aligned}
    \mathcal O_\mathrm{FD}
    =
    &\log_2\!\left(1+\rho G_\mathrm{fwd}(g_\mathrm{DL})\right)
    \\
    \quad+
    &\log_2\!\left(
    1+
    \frac{G_\mathrm{rev}(g_\mathrm{UL})}
    {G_\mathrm{rev}(g_\mathrm{DL})+\rho^{-1}}
    \right),
    \end{aligned}
    \label{eq:O_FD}
    \end{equation}
    where \(\rho=P_\mathrm{T}/\sigma^2\) denotes a common reference SNR for the DL and UL links. Under our assumptions, the DL term is noise-limited, whereas the UL term includes equal-power DMA-side co-channel UL interference from the DL-user direction \(g_\mathrm{DL}\).
\end{itemize}

\textit{Problem Statement:}
For given user locations \(g_\mathrm{DL}\) and \(g_\mathrm{UL}\), and for a given \(\rho\), our optimization problem is summarized as
\begin{equation}
\begin{aligned}
\max_{\substack{
\mathbf v\in\{0,1\}^{N_\mathrm M}\\
\mathbf x\in\mathbb C^{N_\mathrm F-3},\,\|\mathbf x\|_2^2=1\\
\boldsymbol\pi\in\Pi,\; d\in\mathcal D
}}
\mathcal O_\ell(\mathbf v,\mathbf x,\boldsymbol\pi,d), \quad \ell\in\{\mathrm{BF},\mathrm{CH},\mathrm{FD}\}.
\end{aligned}
\label{eq:generic_optimization}
\end{equation}

\textit{Benchmarks:}
We consider two benchmarks. \textit{First,} we define a \textit{reciprocal-DMA benchmark} for the FD objective \(\mathcal O_\mathrm{FD}\). In this case, we replace the circulator by ideal open-circuit terminations. In \eqref{eq_open_circuit_static_reduction}, this corresponds to replacing \(\mathbf S^\mathrm{C}\) by \(\mathbf I_3\). \textit{Second,} to assess the importance of mutual-coupling (MC) awareness, we also consider an \textit{MC-unaware benchmark} for all objectives by assuming \(\bar{\mathbf S}_{\mathcal M\mathcal M}=\mathbf 0\) during optimization. 

\subsection{Optimization Algorithm}

The problem in \eqref{eq:generic_optimization} is mixed discrete-continuous: the DMA control vector \(\mathbf v\) is binary, the circulator-port assignment \(\boldsymbol\pi\) and circulation direction \(d\) are discrete, and the common analog feed-weight vector \(\mathbf x\) is continuous. Our approach for solving \eqref{eq:generic_optimization} is summarized in Algorithm~\ref{alg:greedy_nrdma}. We handle \((\boldsymbol\pi,d)\) by exhaustive search over \(\Pi\times\mathcal D\). For each \((\boldsymbol\pi,d)\), we optimize \(\mathbf v\) by multi-start greedy bit-flip coordinate ascent, which is a simple and effective strategy for discrete RIS-type optimization~\cite{hammami2026statistical}. Since each bit flip changes only one diagonal entry of \(\mathbf\Phi(\mathbf v)\), the matrix inverse in the MNT model admits efficient rank-one Woodbury updates~\cite{prod2023efficient}. For each candidate \(\mathbf v\), we assign \(\mathbf x\) using the closed-form matched-filter heuristic described next.

For a fixed \((\boldsymbol\pi,d,\mathbf v)\), optimizing \(\mathbf{x}\) subject to \(\|\mathbf x\|_2=1\) is generally non-convex. We therefore use a simple matched-filter heuristic. Let
\(\mathbf h_\mathrm{DL}^\top
=\mathbf e_{g_\mathrm{DL}}^\top
\bar{\mathbf T}_\mathrm{fwd}^{(\boldsymbol\pi,d)}(\mathbf r(\mathbf v))\)
denote the effective forward channel from the BS toward the DL user, and let
\(\mathbf h_\mathrm{UL}
=\bar{\mathbf T}_\mathrm{rev}^{(\boldsymbol\pi,d)}(\mathbf r(\mathbf v))
\mathbf e_{g_\mathrm{UL}}\)
denote the effective reverse channel from the UL user to the BS. We set
\begin{equation}
\mathbf x
=
\frac{\mathbf x_\mathrm{DL}+\gamma\mathbf x_\mathrm{UL}}
{\|\mathbf x_\mathrm{DL}+\gamma\mathbf x_\mathrm{UL}\|_2},
\label{eq:x_matched_update}
\end{equation}
where
\(\mathbf x_\mathrm{DL}=\mathbf h_\mathrm{DL}^*/\|\mathbf h_\mathrm{DL}\|_2\)
and
\(\mathbf x_\mathrm{UL}=\mathbf h_\mathrm{UL}^*/\|\mathbf h_\mathrm{UL}\|_2\)
are the matched-filter analog feed-weight directions for the desired DL and UL links, respectively, and \(\gamma\geq0\) controls their relative emphasis.

One sweep of the coordinate ascent on \(\mathbf v\) tests all \(N_\mathrm M\) binary coordinates once. For each coordinate, the corresponding bit is flipped tentatively, the analog feed-weight vector \(\mathbf x\) is recomputed from \eqref{eq:x_matched_update}, and the flip is accepted only if the objective increases. For each random initialization, the search stops when either \(N_\mathrm{sweep}\) sweeps have been completed or a complete sweep accepts no bit flip.

\section{Results}
\label{sec_Results}

\subsection{DMA Prototype and Setup}
\label{subsec_DMA}

\begin{algorithm}
\caption{Greedy NR-DMA optimization}
\label{alg:greedy_nrdma}
\footnotesize
\begin{algorithmic}[1]
\REQUIRE Objective \(\mathcal O_\ell\), sets \(\Pi\), \(\mathcal D\), \(N_\mathrm{init}\), \(N_\mathrm{sweep}\).
\STATE Initialize the best-so-far value as \(\mathcal O^\star\leftarrow-\infty\).
\FORALL{\((\boldsymbol\pi,d)\in\Pi\times\mathcal D\)}
    \STATE Construct \(\bar{\mathbf S}^{(\boldsymbol\pi,d)}\).
    \FOR{\(q=1,\ldots,N_\mathrm{init}\)}
        \STATE Draw random \(\mathbf v\in\{0,1\}^{N_\mathrm M}\).
        \STATE Compute \(\mathbf x\) from \eqref{eq:x_matched_update} and set \(\mathcal O_\mathrm{curr}\leftarrow \mathcal O_\ell(\mathbf v,\mathbf x,\boldsymbol\pi,d)\).
        \FOR{\(s=1,\ldots,N_\mathrm{sweep}\)}
            \STATE Set \(N_\mathrm{acc}\leftarrow0\).
            \FOR{\(m=1,\ldots,N_\mathrm M\)}
                \STATE Form \(\mathbf v'\) by flipping only the \(m\)th entry of \(\mathbf v\).
                \STATE Compute \(\mathbf x'\) from \eqref{eq:x_matched_update} and set \(\mathcal O'\leftarrow \mathcal O_\ell(\mathbf v',\mathbf x',\boldsymbol\pi,d)\).
                \IF{\(\mathcal O'>\mathcal O_\mathrm{curr}\)}
                    \STATE \(\mathbf v\leftarrow\mathbf v'\), \(\mathbf x\leftarrow\mathbf x'\), \(\mathcal O_\mathrm{curr}\leftarrow\mathcal O'\), and \(N_\mathrm{acc}\leftarrow N_\mathrm{acc}+1\).
                \ENDIF
            \ENDFOR
            \IF{\(N_\mathrm{acc}=0\)}
                \STATE \textbf{break}
            \ENDIF
        \ENDFOR
        \IF{\(\mathcal O_\mathrm{curr}>\mathcal O^\star\)}
            \STATE Store \((\mathbf v^\star,\mathbf x^\star,\boldsymbol\pi^\star,d^\star)\leftarrow(\mathbf v,\mathbf x,\boldsymbol\pi,d)\) and set \(\mathcal O^\star\leftarrow\mathcal O_\mathrm{curr}\).
        \ENDIF
    \ENDFOR
\ENDFOR
\end{algorithmic}
\end{algorithm}

Our DMA prototype is displayed in the top panel in Fig.~\ref{fig:multiport_model} and comprises a thin quasi-2D chaotic cavity connected to \(N_\mathrm{F}=7\) feed ports on its back side and patterned with $N_\mathrm{M}=96$ programmable meta-elements on its front side. The cavity mediates strong all-to-all coupling between all feeds and meta-elements, which boosts the DMA's wave-domain flexibility~\cite{prod2025mutual,prod2025benefits} and, in the present work, also amplifies the system-level impact of the circulator-induced non-reciprocity. Each meta-element consists of a complementary electric-LC (cELC) resonator loaded with a PIN diode. The DMA architecture follows that in~\cite{sleasman2020implementation}, except that our prototype uses multiple feeds. We operate the DMA at \(19~\mathrm{GHz}\) and sample its radiated field with a VAA realized by mechanically scanning an open-ended waveguide probe over an \(11\times 11\) square grid. The grid has a spacing of \(2\,\mathrm{cm}\) and is located \(35\,\mathrm{cm}\) in front of the DMA aperture.

The MNT parameters of the setup are not known a priori and cannot be obtained directly. Full-wave simulation is impractical because the DMA is electrically large, particularly sensitive to fabrication tolerances due to its strong MC, and its design details may in principle be proprietary. Direct experimental measurements of the MNT parameters are also impossible because the ``virtual'' ports are not connectorized and too numerous by orders of magnitude for a direct measurement with a conventional vector network analyzer. We therefore rely on the indirect experimental estimation procedure of~\cite{tapie2026experimental}.
This yields a set of proxy MNT parameters that accurately predicts, for arbitrary control vectors \(\mathbf v\), both \(\mathbf T_\mathrm{fwd}(\mathbf v)\) and the feed-port reflection matrix \(\mathbf R(\mathbf v)\). The term ``proxy'' emphasizes that the estimated MNT parameters are not unique, but the inevitable ambiguities are operationally irrelevant because they do not affect the observable mappings \(\mathbf v\mapsto\mathbf T_\mathrm{fwd}(\mathbf v)\) and \(\mathbf v\mapsto\mathbf R(\mathbf v)\). For the present DMA prototype, the corresponding prediction accuracies (see definitions in~\cite[(18)]{tapie2026experimental}) are \(\zeta_{\mathrm{T}_\mathrm{fwd}}=37.7~\mathrm{dB}\) and \(\zeta_\mathrm{R}=40.3~\mathrm{dB}\), respectively.

To construct the NR-DMA's MNT model, we combine these proxy MNT parameters with an experimental measurement of \(\mathbf S^\mathrm{C}\) at 19~GHz for a commercial PE8406 circulator that yields
\begin{equation}
\mathbf S^{\mathrm{C}}
=
\scalebox{0.85}{$
\begin{bmatrix}
0.00-0.03\jmath & 0.04-0.03\jmath & 0.70-0.67\jmath \\
0.96-0.16\jmath & 0.00+0.01\jmath & -0.02-0.03\jmath \\
0.01-0.01\jmath & 0.64-0.73\jmath & -0.02+0.03\jmath
\end{bmatrix}
$}.
\label{eq:S_circ_measured}
\end{equation}
We assume that the circulator is connected to the DMA feeds without additional phase delay or attenuation. We note that knowing $\mathbf{S}_\mathcal{GG}$ is not required to evaluate $\bar{\mathbf T}_\mathrm{fwd}(\mathbf r)$ and $\bar{\mathbf T}_\mathrm{rev}(\mathbf r)$.

We use \(N_\mathrm{init}=12\), \(N_\mathrm{sweep}=20\), and \(\gamma=1\). We randomly draw \(20\) \((g_\mathrm{DL},g_\mathrm{UL})\) pairs on the VAA grid, subject to a minimum grid distance of five. For each SNR value and each user pair, we independently perform the optimization.

\begin{figure}
    \centering
    \includegraphics[width=\linewidth]{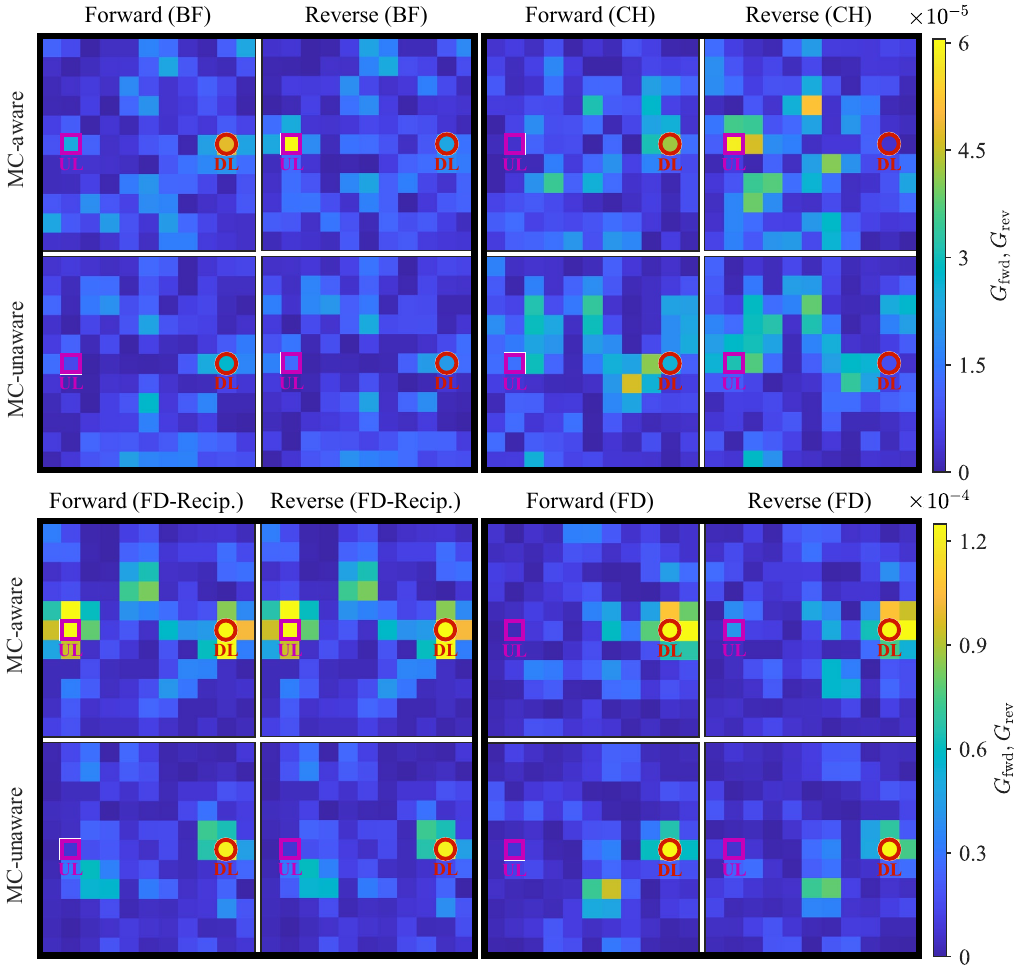}
    \caption{Optimized forward and reverse channel-gain maps with optimized settings for a selected ($g_\mathrm{DL}$,$g_\mathrm{UL}$) pair, for the considered objectives and the reciprocal-DMA benchmark, with and without MC-aware optimization.}
    \label{fig:beamforming}
\end{figure}

\subsection{FD Performance Analysis}
\label{subsec_PerformanceAnalysis}

In Fig.~\ref{fig:beamforming}, we display optimized forward and reverse channel-gain maps for a selected DL/UL user pair. The NR-DMA produces visibly different forward and reverse responses, whereas the reciprocal benchmark yields identical maps in both directions, as expected. The BF objective enhances the reverse gain near the UL marker and the forward gain near the DL marker, while suppressing the opposite user directions. The CH objective shows similar asymmetric channel shaping, but with weaker contrast against the VAA-grid background. The FD objective yields less visually intuitive NR-DMA gain maps because it directly optimizes the sum-rate expression rather than a spatial-contrast metric. For the reciprocal benchmark, the FD objective reduces to a symmetric dual-beamforming pattern. Across all objectives, MC-unaware optimization degrades the realized gain maps, highlighting the need for MC-aware modeling during optimization, in line with~\cite{tapie2026experimental}.

\begin{figure}[!t]
    \centering
    \includegraphics[width=0.8\linewidth]{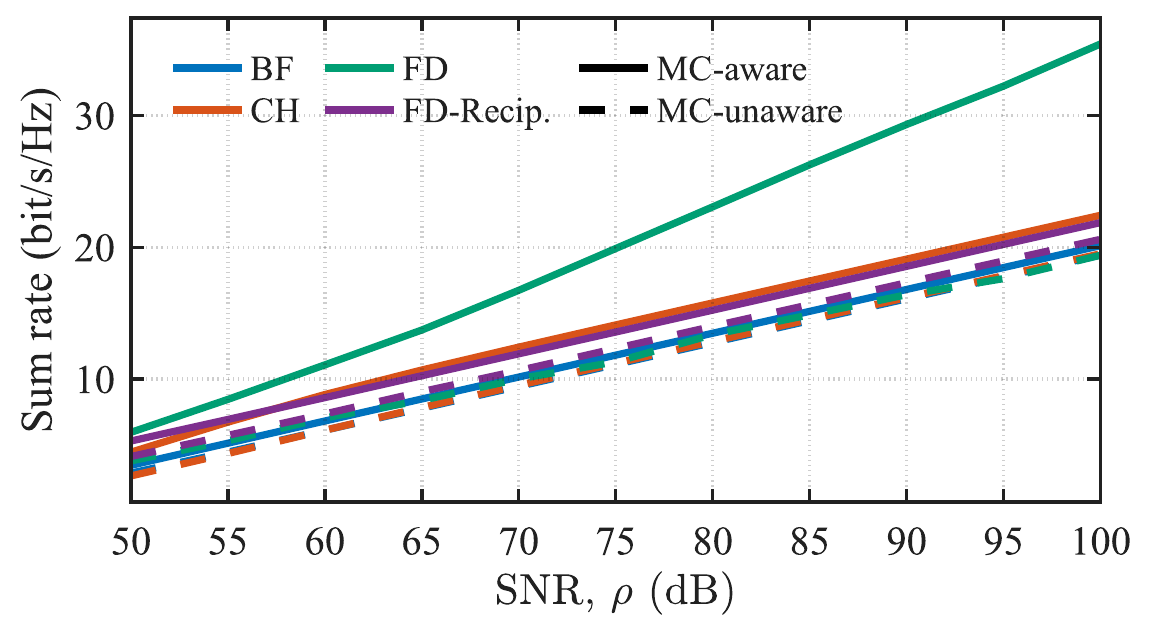}
    \caption{FD sum rate vs. SNR with optimized DMA configurations, averaged over 20 ($g_\mathrm{DL}$,$g_\mathrm{UL}$) pairs, for the considered objectives and the reciprocal-DMA benchmark, with and without MC-aware optimization.}
    \label{fig:capacity}
\end{figure}

In Fig.~\ref{fig:capacity}, we compare achieved sum rates with the eight optimized settings, averaged over the 20 \((g_\mathrm{DL},g_\mathrm{UL})\) pairs. In the high-SNR regime, where multi-user interference is the limiting factor, the MC-aware FD-optimized NR-DMA strongly outperforms the reciprocal benchmark. At $\rho=100 \ \mathrm{dB}$, the NR-DMA and reciprocal benchmark achieve sum rates of \(35.4\) and \(21.9~\mathrm{bit/s/Hz}\), respectively. The proxy objectives perform poorly: the CH-optimized NR-DMA's performance is comparable to the reciprocal benchmark, while the BF-optimized NR-DMA's performance is below the reciprocal benchmark. These observations highlight the importance of end-to-end optimization rather than reliance on proxy objectives. Without MC awareness during optimization, the FD-optimized NR-DMA's performance drops below the reciprocal-DMA benchmark, highlighting again the importance of MC awareness.

\section{Conclusion}
\label{sec_Conclusion}

To summarize, we introduced a practical NR-DMA architecture that adds non-reciprocity to a multi-feed DMA with strong MC by connecting a single circulator to three feed ports. We developed an experimentally grounded MNT model combining proxy DMA parameters of a fabricated DMA prototype with a measured circulator response, and used it to jointly optimize the DMA state, analog feed weights, circulator-port assignment, and circulation direction for FD multi-user operation. Our results show that the NR-DMA can realize clearly distinct forward and reverse channel responses and, in our interference-limited high-SNR case study, substantially outperform a reciprocal DMA benchmark; at \(\rho=100~\mathrm{dB}\), the MC-aware FD-optimized NR-DMA improves on the reciprocal benchmark by roughly \(60\%\). Our results also emphasize the importance of optimizing the final FD objective and accounting for MC.

\bibliographystyle{IEEEtran}

\begin{thebibliography}{10}
\providecommand{\url}[1]{#1}
\csname url@rmstyle\endcsname
\providecommand{\newblock}{\relax}
\providecommand{\bibinfo}[2]{#2}
\providecommand\BIBentrySTDinterwordspacing{\spaceskip=0pt\relax}
\providecommand\BIBentryALTinterwordstretchfactor{4}
\providecommand\BIBentryALTinterwordspacing{\spaceskip=\fontdimen2\font plus
\BIBentryALTinterwordstretchfactor\fontdimen3\font minus \fontdimen4\font\relax}
\providecommand\BIBforeignlanguage[2]{{%
\expandafter\ifx\csname l@#1\endcsname\relax
\typeout{** WARNING: IEEEtran.bst: No hyphenation pattern has been}%
\typeout{** loaded for the language `#1'. Using the pattern for}%
\typeout{** the default language instead.}%
\else
\language=\csname l@#1\endcsname
\fi
#2}}

\bibitem{shlezinger2021dynamic}
N.~Shlezinger \emph{et~al.}, ``Dynamic metasurface antennas for {6G} extreme massive {MIMO} communications,'' \emph{IEEE Wirel. Commun.}, vol.~28, no.~2, pp. 106--113, 2021.

\bibitem{yuan2026single}
S.~S.~A. Yuan \emph{et~al.}, ``Single-frequency symmetry-empowered through-barrier sensing in reconfigurable complex media,'' \emph{arXiv:2606.05877}, 2026.

\bibitem{prod2025mutual}
H.~Prod'homme and P.~del Hougne, ``Mutual coupling in dynamic metasurface antennas: Foe, but also friend,'' \emph{IEEE Wirel. Commun.}, vol.~32, no.~4, pp. 30--36, 2025.

\bibitem{prod2025benefits}
H.~Prod'homme \emph{et~al.}, ``Benefits of mutual coupling in dynamic metasurface antennas,'' \emph{IEEE Trans. Antennas Propag.}, vol.~74, no.~3, pp. 2589--2604, 2025.

\bibitem{wang2024channel}
H.~Wang \emph{et~al.}, ``Channel reciprocity attacks using intelligent surfaces with non-diagonal phase shifts,'' \emph{IEEE Open J. Commun. Soc.}, vol.~5, pp. 1469--1485, 2024.

\bibitem{xu2025non}
J.~Xu \emph{et~al.}, ``Non-reciprocal reconfigurable intelligent surfaces,'' \emph{IEEE Wirel. Commun. Lett.}, vol.~14, no.~10, pp. 3329--3333, 2025.

\bibitem{li2025non}
H.~Li and B.~Clerckx, ``Non-reciprocal beyond diagonal {RIS}: Multiport network models and performance benefits in full-duplex systems,'' \emph{IEEE Trans. Commun.}, vol.~73, no.~11, pp. 12,221--12,234, 2025.

\bibitem{liu2026non}
Z.~Liu \emph{et~al.}, ``Non-reciprocal beyond diagonal {RIS}: Sum-rate maximization in full-duplex communications,'' \emph{IEEE Trans. Commun.}, vol.~74, pp. 5169--5183, 2026.

\bibitem{del2025physics}
P.~del Hougne, ``A physics-compliant diagonal representation for wireless channels parametrized by beyond-diagonal reconfigurable intelligent surfaces,'' \emph{IEEE Trans. Wirel. Commun.}, vol.~24, no.~7, pp. 5871--5884, 2025.

\bibitem{prod2025beyond}
H.~Prod'homme and P.~del Hougne, ``Beyond-diagonal dynamic metasurface antenna,'' \emph{IEEE Commun. Lett.}, vol.~30, pp. 258--262, 2025.

\bibitem{tapie2026experimental}
J.~Tapie and P.~del Hougne, ``Experimental multiport-network parameter estimation for a dynamic metasurface antenna,'' \emph{IEEE Trans. Antennas Propag.}, 2026.

\bibitem{hammami2026statistical}
C.~Hammami \emph{et~al.}, ``Statistical multiport-network modeling and efficient discrete optimization of {RIS},'' \emph{IEEE Wirel. Commun. Lett.}, vol.~15, pp. 1425--1429, 2026.

\bibitem{prod2023efficient}
H.~Prod'homme and P.~del Hougne, ``Efficient computation of physics-compliant channel realizations for (rich-scattering) {RIS}-parametrized radio environments,'' \emph{IEEE Commun. Lett.}, vol.~27, no.~12, pp. 3375--3379, 2023.

\bibitem{sleasman2020implementation}
T.~A. Sleasman \emph{et~al.}, ``Implementation and characterization of a two-dimensional printed circuit dynamic metasurface aperture for computational microwave imaging,'' \emph{IEEE Trans. Antennas Propag.}, vol.~69, no.~4, pp. 2151--2164, 2020.

\end{thebibliography}

\end{document}